\documentclass[aps,%prd,
groupedaddress,showpacs,nofootinbib,prl,twocolumn]{revtex4}
\usepackage{graphicx,amssymb,amsmath,epsfig}
\usepackage[english]{babel}
\usepackage{graphicx}% Include figure files
\usepackage{dcolumn}% Align table columns on decimal point
\usepackage{bm}% bold mathe

\def\comment#1{}
\def\IncludeEpsImg#1#2#3#4{\renewcommand{\epsfsize}[2]{#3##1}{\epsfbox{#4}}}

\def\Omegabf{{\boldsymbol\Omega}}
\def\nablabf{{\boldsymbol\nabla}}
\def\deltabf{{\boldsymbol\delta}}

\newcommand{\be}{\begin{equation}}\newcommand{\ee}{\end{equation}}
\newcommand{\bea}{\begin{eqnarray}}\newcommand{\eea}{\end{eqnarray}}
\newcommand{\beaa}{\begin{eqnarray}}\newcommand{\eeaa}{\end{eqnarray}}
\newcommand{\ba}{\begin{array}}\newcommand{\ea}{\end{array}}
\newcommand{\bit}{\begin{itemize}}\newcommand{\eit}{\end{itemize}}
\newcommand{\ben}{\begin{enumerate}}\newcommand{\een}{\end{enumerate}}

 \newcommand{\sfrac}[2]{\raisebox{0.095ex}{\scriptsize${\frac{#1}{#2}}$}}
 
 \newcommand{\sbf}[1]{\mbox{\scriptsize\bf{#1}}}

\def\lfrac#1#2{#1/#2}

\begin{document}

\title{New Gauge Symmetry in Gravity and the Evanescent Role of Torsion
}

\author{H. Kleinert}

%\vspace{2mm}

       \address{
Institut f\"ur Theoretische Physik,
 Freie Universit\"at Berlin,
Arnimallee 14, D14195 Berlin, Germany\\
ICRANeT, Piazzale della Republica 1, 10 -65122, Pescara, Italy
}

%\maketitle

\vspace{2mm}

\begin{abstract}

If the Einstein-Hilbert action  ${\cal L}_{\rm EH}\propto R$ is re-expressed in
Riemann-Cartan spacetime using the gauge fields of translations, the
vierbein field  $h^\alpha{}_\mu$,
and the gauge field of local Lorentz transformations, the spin connection
$A_{\mu \alpha }{}^ \beta $, there exists a new gauge symmetry
which permits reshuffling the torsion, partially or totally,
into the Cartan curvature term of the Einstein tensor, and back, via a
{\em new multivalued gauge transformation\/}. Torsion can be chosen at will
by an arbitrary gauge fixing functional. There exist many equivalent ways of
specifying the theory, for instance Einstein's traditional way where
${\cal L}_{\rm EH}$ is expressed completely  in terms of the metric
$g_{\mu \nu }=h^ \alpha {}_\mu h_ \alpha {}_ \nu $, and the torsion is zero,
or Einstein's  teleparallel formulation, where ${\cal L}_{\rm EH}$  is expressed
in terms of the torsion tensor, or an infinity of  intermediate ways.
As far as the gravitational field in the far-zone of a celestial object
is concerned, matter composed of spinning particles can be replaced by matter
with only orbital angular momentum, without changing the long-distance forces,
no matter which of the various new gauge representations is used.

\end{abstract}


\maketitle

{\bf 1.}
 In theoretical physics
it often happens that
a mathematical structure has
a simple extension for which
a
natural phenomenon
is waiting to be discovered.
The most prominent example
is the existence
of a negative square root
of the relativistic mass shell relation
$p_0= \sqrt{{\bf p}^2+m^2}$
which led Dirac to postulate
the existence of a positron,
discovered in 1932 by Carl Anderson \cite{AND}.
Sometimes, this
rule does
not seem to work
initially, only to find out
later
that
nature has chosen an unexpected way to
make it work after all.
Here the best example
is
the existence
of a solution of the above energy-momentum
relation for negative
$m^2$, which was
interpreted by some theoreticians
as the signal for the existence of a
particle faster than light. Such particles were never found.
A simple
physical realization appeared, however,
with the discovery
of the Ginzburg-Landau field theory
of phase transitions and its quantum versions
(now referred to as Higgs field theory).
Since there are
always
{\em interactions\/},
a negative parameter $m^2$
destabilizes
the field fluctuations.
The fields
move to a new ground state, around
which they fluctuate
with {\em positive\/} $m^2$.
The situation is completely analogous
to what happens in any building
if $ \omega ^2$
of
one of its eigenfrequencies
turns negative.
The building collapses until the debris
settles in a ruin, and that has only positive $ \omega ^2$'s.
The collapse
of an
interacting field system
with
negative $m^2$
is observed as
a phase transition to
a state
with
stable fluctuations
and positive
 $m^2$.\\[-1.2em]

{\bf 2.}
For many years, theorists have been wondering
why Einstein's theory of gravity
represents such a perfect
geometrization of the gravitational forces \cite{WB}.
 Since the work of Cartan in 1922 it is known that
the Riemannian spacetime,
in which the
celestial objects
move,
has a ``natural"
extension
to {\em Riemann-Cartan\/}
spacetime. This
 possesses
a further geometric property
called {\em torsion\/}.
Why is there no trace of it in the movements of planets?
Einstein himself has
asked
this question
and discussed it in
letters
with
Cartan \cite{EINSC}.
He
set up a theory of {\em teleparallelism\/}
which
explains gravity
by a theory in Riemann-Cartan
spacetime,
in which the total curvature tensor
vanishes identically.
The Einstein-Hilbert action is then
 equal to a combination of scalars
formed from torsion tensors
\cite{HS}, and torsion forces
provide us merely with an alternative way of
describing gravitational forces, as emphasized in Refs.~\cite{AP,PER}.

{\bf 3.}
 Yet another extension of
Einstein's theory
to Riemann-Cartan spacetime
was advanced
since
1959
 \cite{all,GFCM,MVF}.
It
has
 the appealing feature that it can be
rewritten
as a gauge theory
invariant under local Poincar\'e 
transformations, i.e.,
both local translations and local Lorentz
transformations, thus
bringing it to a similar form
as the gauge theories
of weak, electromagnetic, and strong interactions.
This gauge theory treats torsion as an {\em independent\/} 
field which couples only to the intrinsic spin of 
the elementary particles in a celestial body.
Unfortunately, however, such an approach has several
unsatisfactory features.
First, the theory is meant to be classical, but the
spin carries a power of $\hbar$ which vanishes
in the classical limit. So there is really no classical source 
of torsion.
Indeed, if torsion 
couples to spin with the 
gravitational 
coupling strength, the factor $\hbar$
implies that it cannot play any sizable role in
the forces between celestial bodies.
For example, even if the earth consisted only of polarized  
atoms, its intrinsic spin would be $10^{-15}$ times smaller than the 
rotational spin 
around the axis.

\begin{figure}[ht]
\begin{picture}(105,40)
\put(0,-0){\IncludeEpsImg{105.64mm}{26.43mm}{.10000}{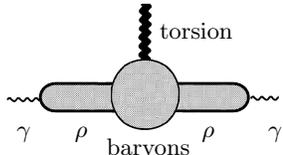}}
\put(5,-2){$\gamma$}
\put(100,-2){$\gamma$}
\put(39,-9){baryons}
\put(27,-2){$\rho$}
\put(75,-2){$\rho$}
\put(59,35){torsion}
\end{picture}
	\caption{Diagram for mass generation of photon.
It couples via a $\rho$-meson  
to baryonic matter
which would be coupled to torsion if $q\neq 1$.}
	\label{@baryon}
\end{figure}

Moreover, there exist severe conceptual problems.
One was emphasized in Ref.~\cite{OA}. As long as we do not know precisely the truly {\em elementary
  particles\/}, and it is doubtful that we ever will,
many particles are described by {\rm effective fields\/}, and it is impossible
to specify whether the spin of those fields is caused by orbital
motion or by the intrinsic spins of more elementary constituents.
As an example, the spin-one field of a $\rho$-meson 
contains a wave function of two spinless pions in a p-wave,
which do not couple to torsion.
But it also contains two spin-$\sfrac{1}{2}$ quarks in an s-wave 
which would couple.
Another problem
is 
that
if torsion couples to all spins, the photon
becomes massive. In order to avoid this,
the authors
advocating this approach postulate that the photon is
is an exception, and is not coupled
to torsion.
However, this contradicts the 
fact that roughly one percent 
of a photon is a virtual 
$\rho$-meson, which 
is strongly coupled to baryons.
These, in turn, 
are supposed to be coupled to torsion (see Fig.~\ref{@baryon}), so
the photon would become massive after all.

Thus the existence of
an independent torsion field
is highly dubious,
\comment{Until the recent results of the
satellite experiment
Gravity Probe B
\cite{STAN}
some people hoped
that the observed Lense-Thirring effect
would deviate from Einstein's prediction.
But it did not.
Thus a simple possiblity
of discovering
a  torsion field has faded \cite{HOWE}.
}%
and we may ask ourselves,
 whether the 
description of gravity
in Riemann-Cartan spacetime
proposed in 
Refs. \cite{all,GFCM,MVF}
has really a chance of being true, or
whether nature
doesn't have a deeper reason
 for avoiding the above problems.
It is the purpose of this note to
answer this question affirmatively.
Inspiration
comes from
a simple model
of gravity, a ``world crystal" with defects \cite{WC,MVF,ENTR}, whose lattice constant
is of the order of a Planck length. %(to be set equal to unity in the sequel).
Some consequences of
 such a world crystal were pointed out
in  a recent study of black holes  in such a scenario \cite{JKS}.

{\bf 4.} We begin by  showing
that in the absence of matter,
a world crystal is a model
for Einstein's theory
with a new type of extra gauge symmetry in which
zero torsion is merely a special gauge.
A completely equivalent gauge
is the absence
of Cartan curvature, which
is found in Einstein's teleparallel universe.
Before presenting the argument,
recall that
a crystal can have two
different types of topological line-like defects \cite{GFCM,MVF},
which in a four-dimensional world crystal are world surfaces (which may be the objects
of
string theory).

\begin{figure}[htb]
\begin{picture}(105.64,39.645)
%\put(0,-14){\IncludeEpsImg{105.64mm}{26.43mm}{.020000}{fig-2-2.eps}}
\put(0,-14){\includegraphics[width=2.9cm]{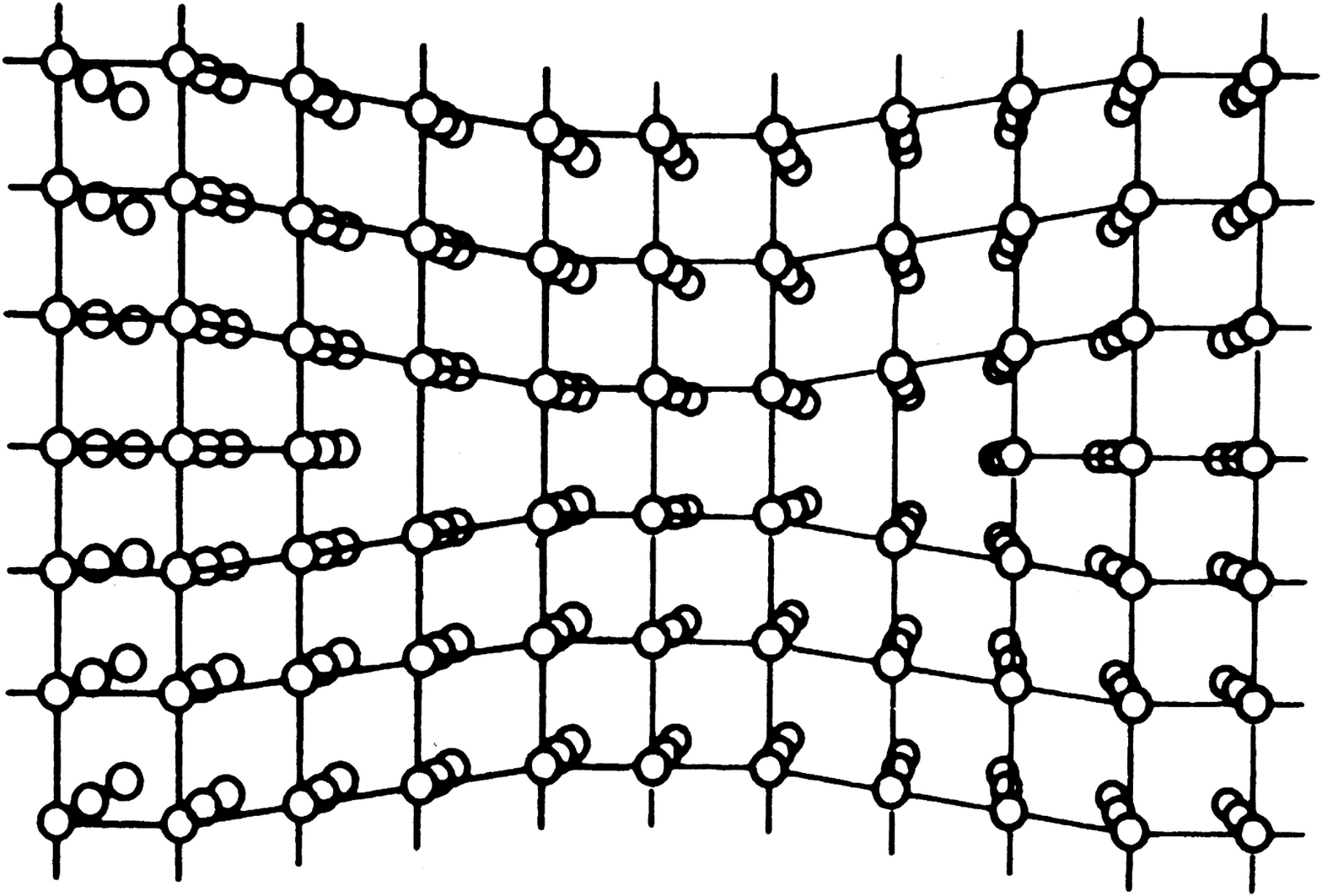} }
\end{picture}
\caption[Formation of a dislocation line (of the edge type) from a disc
of missing atoms]
{Formation of a dislocation line (of the edge type) by
a Volterra process.
The Burgers vector ${\bf b}$ characterizes the missing layer.
There exist two more types where
${\bf b}$ points in orthogonal directions.
}
\label{5Fig. 2.2}\end{figure}

First, there are translational defects
called {\em dislocations\/}
(Fig.~\ref{5Fig. 2.2}).
\comment{
      \begin{figure}[hbtb]
\begin{picture}(105.64,154.645)
\put(0,-8){\IncludeEpsImg{105.64mm}{26.43mm}
{.020000}{fig-2-12.eps}}
\end{picture}
\caption[Three different possibilities of constructing
disclinations]
{Three different possibilities of constructing
disclinations: (a) wedge, (b) splay, and (c) twist disclinations.
They are characterized by the Frank vector $ \Omegabf $.}
\label{5Fig. 2.12.(a-c)}\end{figure}
}%
\begin{figure}[hbtb]
\begin{picture}(105.64,114.645)
\put(-60,72){\IncludeEpsImg{105.64mm}{26.43mm}{.040000}{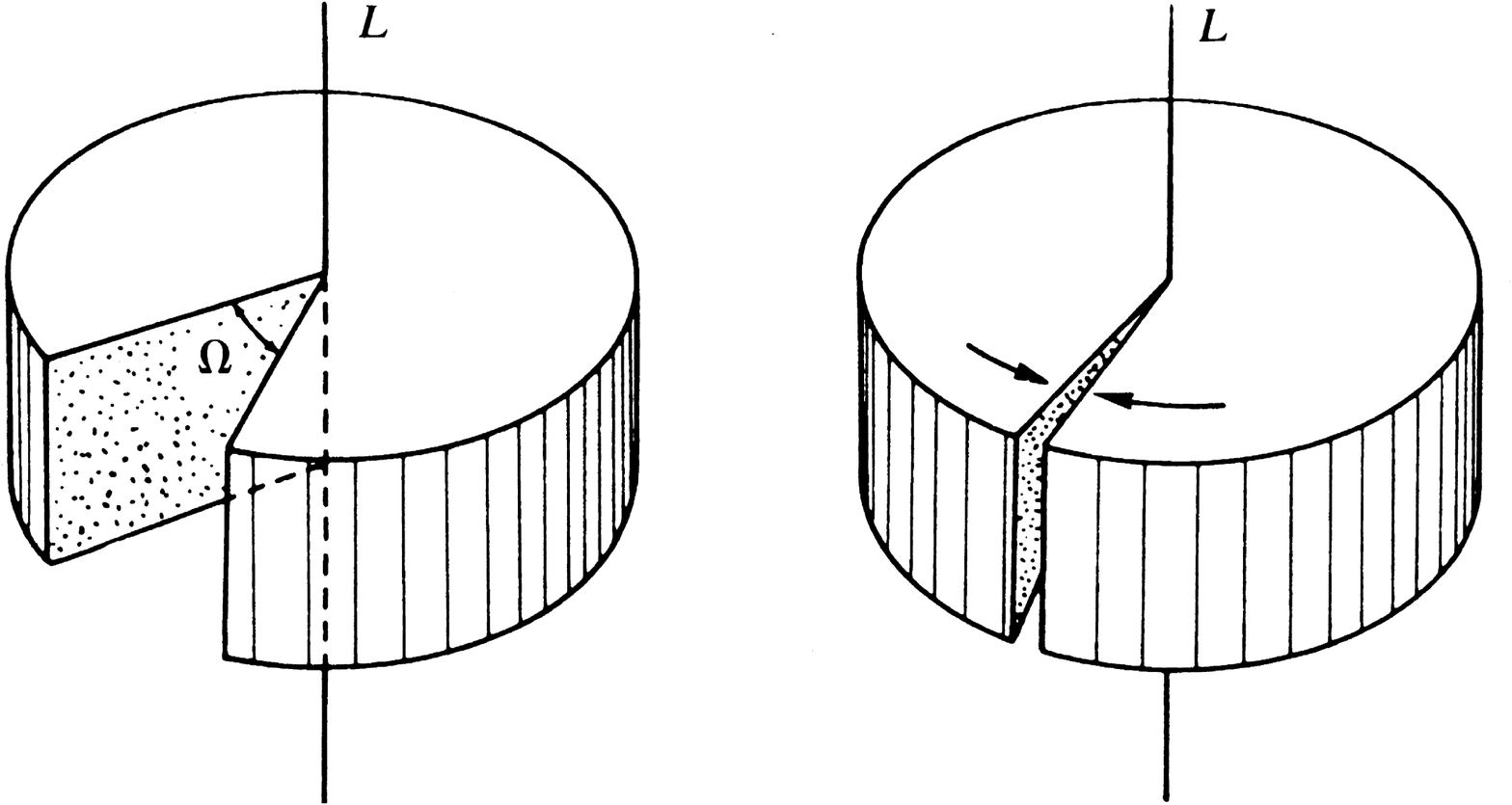}}
\put(10,28){\IncludeEpsImg{105.64mm}{26.43mm}{.040000}{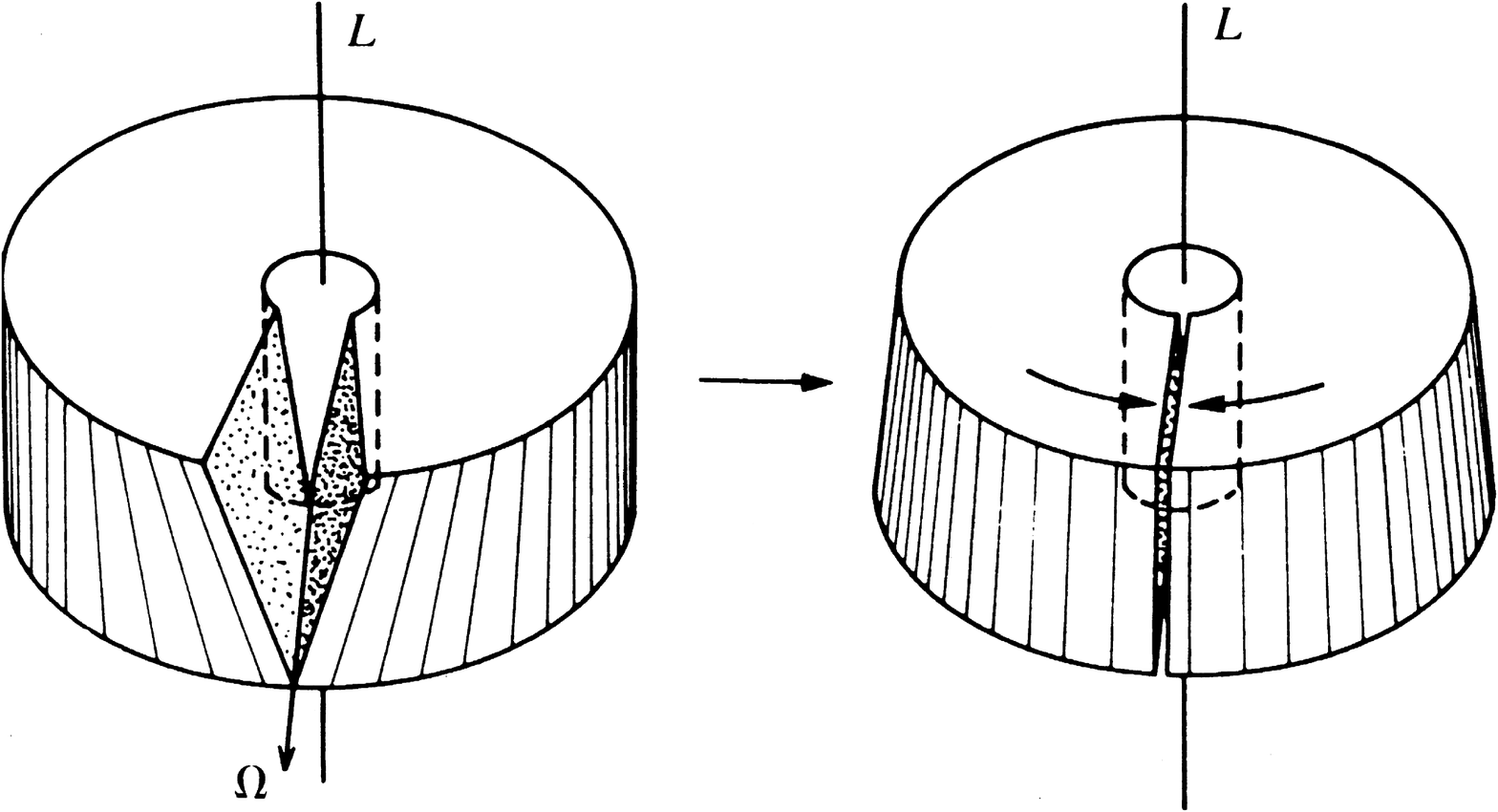}}
\put(70,-10){\IncludeEpsImg{105.64mm}{26.43mm}{.040000}{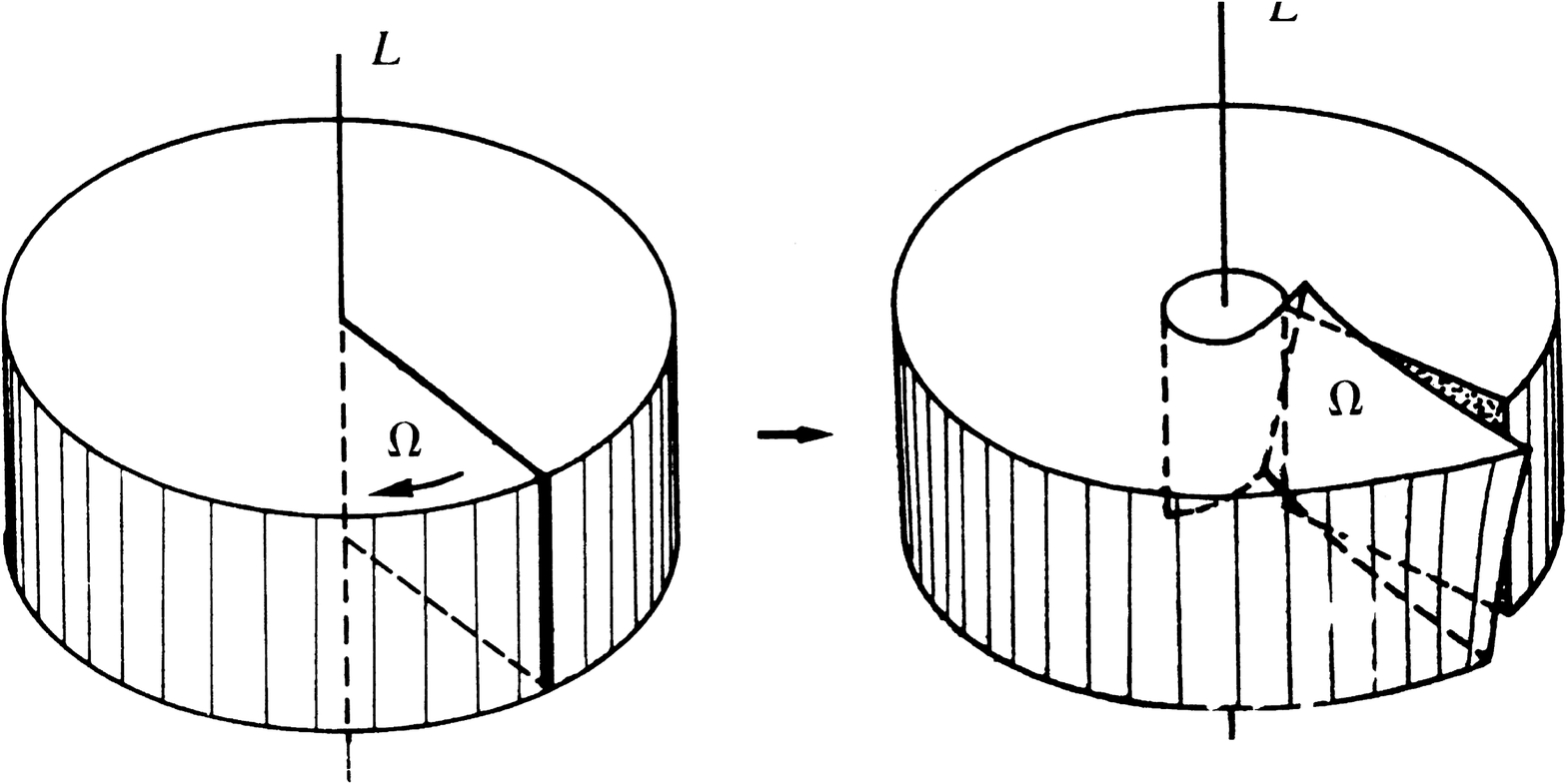}}
\end{picture}
\caption[Three different possibilities of constructing
disclinations]
{Three different possibilities of constructing
disclinations:  wedge,  splay, and twist disclinations.
They are characterized by the Frank vector $ \Omegabf $.}
\label{5Fig. 2.12.(a-c)}\end{figure}%
These are produced
by
a cutting process due
to Volterra: a single-atom  layer
is removed from the crystal, allowing
the remaining atoms to relax
to equilibrium under the elastic forces.
A second type of topological defects
is of the rotation type,
the so-called {\em disclinations\/} (Fig. \ref{5Fig. 2.12.(a-c)}).
They arise by removing an entire wedge
from the crystal and re-gluing the free surfaces.

The defects imply
a failure of derivatives to commute in front of the displacement
field
$u_i({\bf x})$. In three dimensions,
the dislocation
density
is given by the tensor
\vspace{-1.7em}~\\
\begin{equation}
 \alpha _ {ij}({\bf x})= \epsilon _{ikl}
\nabla_k\nabla_l u_j({\bf x}).
\label{@}\end{equation}
\vspace{-1.7em}~\\
 If
$ \omega _i\equiv \frac{1}{2}\epsilon _{ijk}[
\nabla _ju_k({\bf x})
-\nabla _ku_j({\bf x})] $
denotes
the local rotation field,
the disclination density
is defined by
\vspace{-1.2em}~\\
\begin{equation}
\theta_{ij}({\bf x})= \epsilon _{ikl}
\nabla_k\nabla_l  \omega _j({\bf x}).
\label{@}\end{equation}
\vspace{-1.7em}~\\
\vspace{-2.em}~\\

The defect densities satisfy the conservation laws
\vspace{-2.em}~\\
\begin{equation}
\nabla_i \theta _{ij}=0,~~~
\nabla_i \alpha _{ij}=- \epsilon _{jkl}\theta_{kl}.
\label{@CL}\end{equation}
\vspace{-2.em}~\\
These are fulfilled as Bianchi identities
if we express $\theta_{ij}({\bf x})$,
$\alpha_{ij}({\bf x})$
in terms of plastic
gauge fields $\beta ^p_{kl},\phi ^p_{lj}, $
setting
%
%\begin{equation}
$\theta_{ij}= \epsilon _{ikl}\nabla_k \phi ^p_{lj},~
 \alpha _{il}= \epsilon _{ijk}\nabla_j \beta ^p_{kl} +
\delta _{il}\phi^p_{kk}\!-\!\phi^p _{li}$.
%\nonumber \end{equation}
%
The defect densities are invariant under the
gauge transformations
 $ \beta ^p_{kl}\rightarrow
\beta ^p_{kl}+\nabla_k u_l^p- \epsilon _{klr} \omega ^p_r$,\,
$
\phi^p_{li}\rightarrow
\phi^p_{li}+\partial _l \omega ^p_i$,
where
$ \omega ^p_i\equiv \sfrac{1}{2} \epsilon_{ijk} \nabla_j u^p_k$.
Thus
$h_{ij}\equiv
\beta^p_{ij}+ \epsilon _{ijk} \omega ^p_k$ and
$
A_{ijk}\equiv
\phi^p_{ij} \epsilon _{jkl}$ are
{\em translational\/} and {\em rotational\/} defect gauge fields
in the crystal \cite{DEFS}.

The Volterra processes
can be represented mathematically
by multivalued transformations
from an
 Euclidean
 crystal with coordinates
$\bar x^{a}$
to a crystal with defects
and coordinates
$x^{\mu}$,
%The transformation is not integrable
%and can only be specified
%by the transformation
%of the coordinate differentials
as illustrated  in Figs.~\ref{disloc1} and~\ref{disclin1}
for two-dimensional crystals.
{\begin{figure}[htb]
\begin{picture}(105.64,30.64)
%\put(-13,-10){\IncludeEpsImg{105.64mm}{26.43mm}
%{.2000}{disloc1.eps}}
\put(-19,-10){\IncludeEpsImg{105.64mm}{26.43mm}
{.176000}{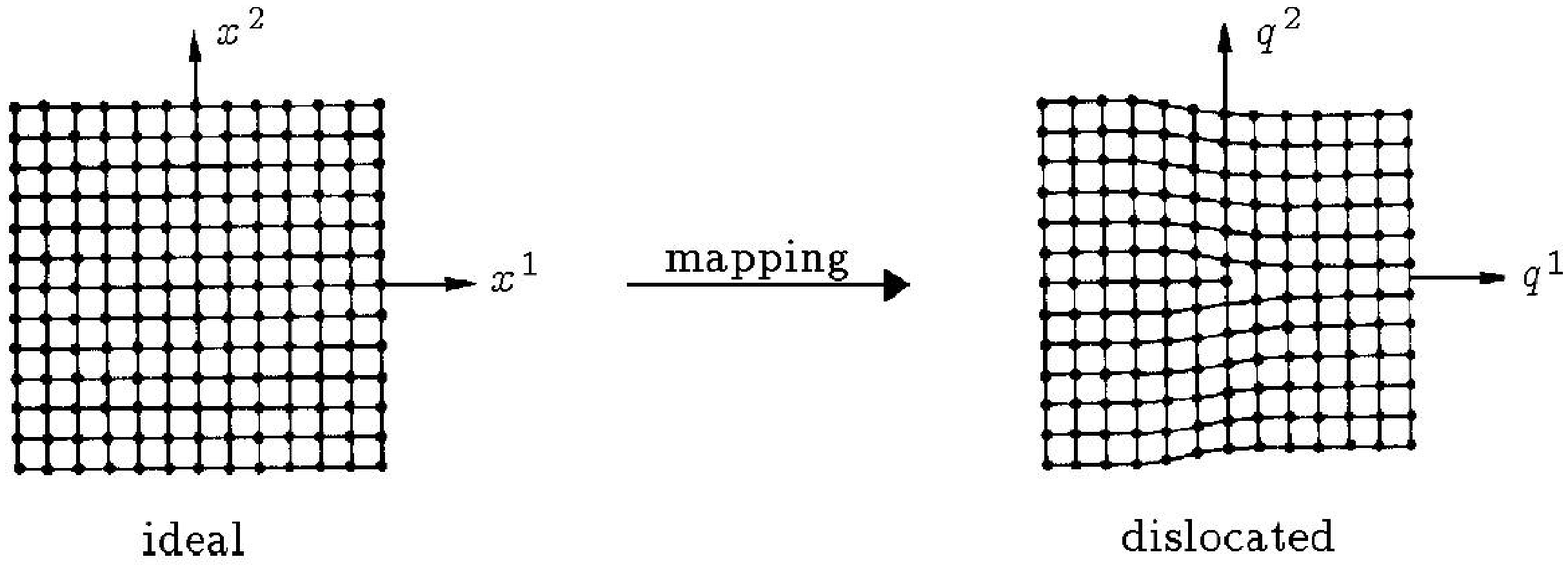}}
\end{picture}
\caption[Multivalued mapping of the perfect
crystal to an edge dislocation
]
{Multivalued mapping of the perfect
crystal to an edge dislocation
with a Burgers vector ${\bf b}$ pointing in the 2-direction.
}
\label{disloc1}\end{figure}
     \begin{figure}[htb]
\begin{picture}(105.64,30.645)
\put(-18,-10){\IncludeEpsImg{105.64mm}{26.43mm}
{.25000}{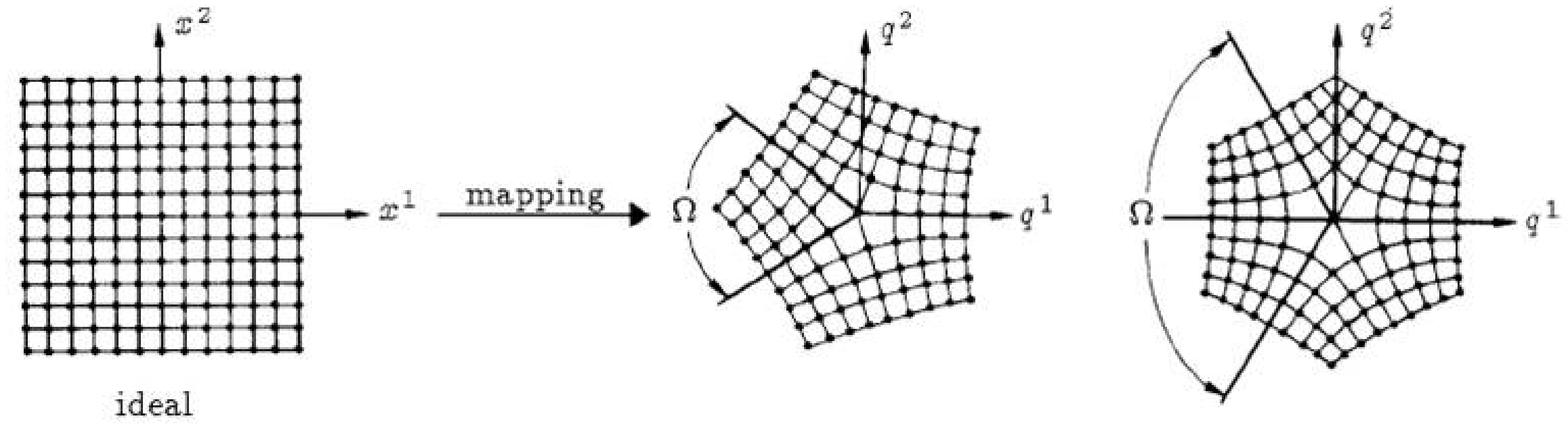}}
\end{picture}
\caption[
Multivalued mapping of the perfect
crystal to a wedge disclination
]
{
Multivalued mapping of the perfect
crystal to a wedge disclination
of Frank vector $ \Omega $ in the third direction.
}
\label{disclin1}\end{figure}
}
\comment{
\begin{figure}[htb]
\begin{picture}(105.64,72.645)
\put(-19,39){\IncludeEpsImg{105.64mm}{26.43mm}
{.176000}{disloc1.eps}}
\put(-18,-10){\IncludeEpsImg{105.64mm}{26.43mm}
{.25000}{disclin1.eps}}
\end{picture}
\caption[Multivalued mapping of the perfect
crystal to an edge dislocation
]
{Multivalued mapping of the perfect
crystal to a crystal with an edge dislocation
%of Burgers vector ${\bf b}$ pointing in the 2-direction,
and  with a wedge disclination.
%of Frank vector $ \Omega $ in the third direction.
}
\label{disclin1}\end{figure}
}%

For an edge dislocation the mapping is
%
%\begin{eqnarray}
$\bar x^1=x^1,~
\bar x^2=x^2+(\lfrac{b}{2\pi})\phi (x)$,
%\label{@}\end{eqnarray}
%
where
$\phi(x)\equiv (1/2\pi)
\arctan (x^2/x^1)$. Initially, this function
has a cut from the origin towards left infinity.
In a second step, the cut is removed and the multivalued version
of the arctan is taken.
This makes $\phi({\bf x})$
the Green function of the commutator
$[\partial _1,\partial _2]$:
%
%\begin{eqnarray}
$(\partial _1\partial_2 -
\partial _2\partial_1)\phi(x)=  \delta^{(2)} ({\bf x})$.
%\label{@}\end{eqnarray}
%
For a wedge disclination, the mapping is
%
%\begin{eqnarray}
$d\bar x^i= \delta ^i{}_\mu
\left[x^\mu+(\lfrac{ \Omega }{2\pi} )\varepsilon ^\mu{}_ \nu x^ \nu \phi(x) \right] $.
%\label{@dl}\end{eqnarray}
%

A combination of the two
~\\[-1.5em]
\begin{eqnarray}
 \!\!\!\!\eta _{ij}({\bf x})\!\!\!&\equiv&\!\!\!
\theta_{ij}({\bf x})
% \nonumber \\ &-&
\!-\!\sfrac{1}{2}
\nabla_m
[
   \epsilon  _{min}
 \alpha _{jn}({\bf x})+\{ij\}
+ \epsilon _{ijn} \alpha _{mn}]
\label{@dc}\end{eqnarray}
~\\[-1.5em]
forms the
 {\em defect tensor\/}
\\\vspace{-1.6em}
\begin{eqnarray}
 \eta _{ij}({\bf x})\equiv  \epsilon _{ikl}
 \epsilon _{jmn}
\nabla_k
\nabla_m u^p_{ln}({\bf x}),~~~u^p_{ln}\equiv \sfrac{1}{2}
(
  \beta ^p_{ln}
+  \beta ^p_{nl})      .
\label{@def}\end{eqnarray}
\vspace{-1.6em}~\\
It is a symmetric tensor
due to the conservation laws
(\ref{@CL}),
and represents the Einstein tensor
$
G_{ij }\equiv
R_{ij }-\sfrac{1}{2}
g_{ij}R_{k }{}^ k
$
of the
 geometry
 of the world crystal.

The expressions
can easily be defined on a simple-cubic
world lattice
if we replace $\nabla_i$ by lattice derivatives, as shown in
\cite{GFCM,MVF}. There it is also
shown
that,
in three spacetime dimensions,
 the disclination
density
$\theta_{ij}({\bf x})$  represents the
Einstein tensor
$G^{{\rm C}}_{ij}$
associated with the
{\em Cartan curvature tensor\/}
$R^{{\rm C}}_{ijk }{}^l$
of the Riemann-Cartan geometry
 of the world crystal.
The relation is
\vspace{-1.8em}~\\
\begin{eqnarray}
G^{{\rm C}}_{ji}({\bf x})
= \epsilon _{ikl}\nabla_k\nabla_l \omega _j({\bf x})
= \theta_{ij}({\bf x})
.
\label{@}\end{eqnarray}
\vspace{-1.9em}~\\
The dislocation density
$\alpha_{ij}({\bf x})$ represents the torsion $S_{lkj}=
\sfrac{1}{2}(
\Gamma _{lkj}-
\Gamma _{klj})
$
of the Riemann-Cartan geometry.
Here the relation is
\vspace{-1.em}
\begin{eqnarray}
%S_{klj}=\sfrac{1}{2} \epsilon _{kli} \alpha _{ij}.
 \alpha _{ij}
= \epsilon _{ikl}
S_{lkj}.
\label{@TOR}\end{eqnarray}
\vspace{-1.9em}

{\bf 5.}
The standard form of
a defect with Burgers vector $b_l$ and Frank vector
$ \Omega _q$ has a
displacement field
\begin{equation}
 u_l     ({\bf x})=- \delta ({\bf x},V)
[b_l+  \epsilon  _{lqr} \Omega _q(x_r-\bar x_r)],
\label{@DISL}\end{equation}
where
$\epsilon _{lqr}$
is the
antisymmetric unit  tensor,
$\bar x_r$
the axis of rotation of the disclination part,
 and $\delta ({\bf x};V)$  is the delta function
on the volume $V$, i.e.,
in three dimensions:
\vspace{-2em}~\\
\begin{eqnarray}
\delta({\bf x};V)=\int_V
d^3 { x}'
\, \delta^{(3)}({\bf x}- {\bf x}').
\label{10@.xDelV}\end{eqnarray}
\vspace{-1.6em}~\\
Its derivative
is the  delta function on the Volterra surface $S$ of $V$:
\vspace{-1.9em}~\\
\begin{eqnarray}
-\nablabf\delta({\bf x};V)=\deltabf({\bf x};S)=\int_S
d {\bf S}'
\, \delta^{(3)}({\bf x}- {\bf x}').
\label{10@.xDel}\end{eqnarray}
\vspace{-2.9em}~\\

\begin{figure}[tbhp]
\begin{picture}(105.64,25.645)
\put(-55,-10){\IncludeEpsImg{105.64mm}{26.43mm}{.050000}{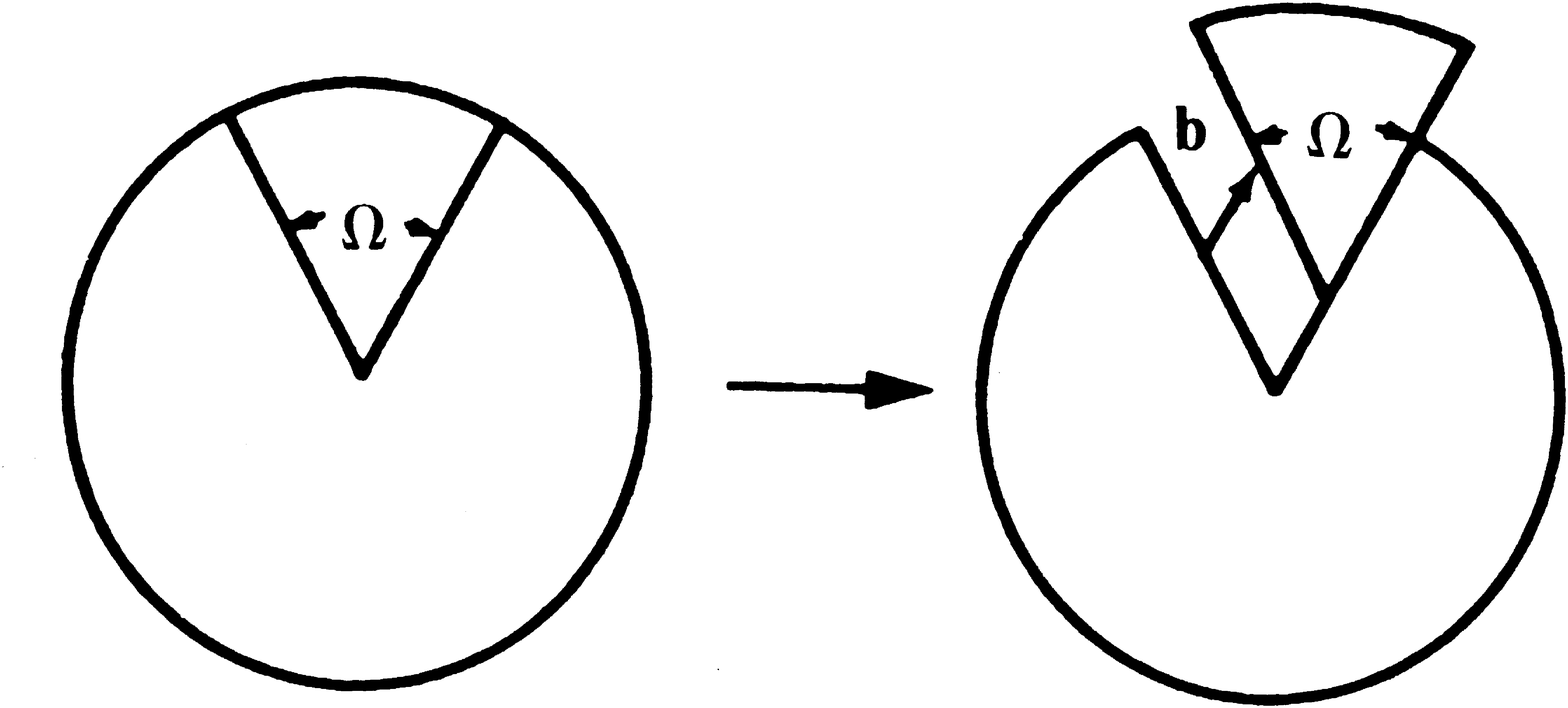}}
\put(55,-5.4){\IncludeEpsImg{105.64mm}{26.43mm}{.315000}{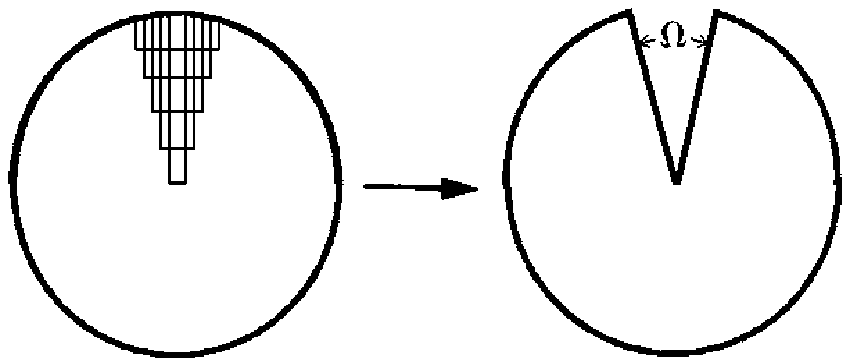}}
\put(-58,-2.4){\tiny$a)$}
\put(52,-2.4){\tiny $b)$}
\end{picture}
\caption[]
{Equivalence between
a)
dislocation and pair of disclination lines,
b) disclination and
stack of dislocation lines.
}
\label{stack}\end{figure}

For the new gauge symmetry, the crucial observation
is that as a simple consequence
of
(\ref{@DISL}), a dislocation line in the world crystal
can either be obtained by  a Volterra process
of cutting out a thin slice of material
of thickness ${\bf b}$,
or alternatively by
cutting out a wedge
of Frank vector $ \Omegabf $, and reinserting it
at distance ${\bf b}$ from the cut.
Thus the
dislocation line
is indistinguishable from a
 pair of disclination lines
with opposite Frank vector $ \Omega $
whose axes of rotation are separated by
a distance ${\bf b}$
(Fig.~\ref{stack}a).  Conversely,
 a
disclination line is equivalent to a stack of dislocation lines
with fixed Burgers vector ${\bf b}$
(Fig.~\ref{stack}b).

\comment{
The equivalence works also in the opposite direction, as seen in Fig.~\ref{stack}b).
The corresponding relation between the plastic displacements
in (\ref{@DISL})
 is obvious.}

Analytically, this is most easily seen in the two-dimensional
version of the relation (\ref{@dc}):
\vspace{-1.8em}~\\
\begin{eqnarray}
 \eta _{33}=\theta_{33}+ \epsilon _{3mn} \nabla_m  \alpha _{3n}.
\label{@2def}\end{eqnarray}
\vspace{-1.8em}~\\
Each term is invariant under
the plastic gauge transformations
 $ \beta ^p_{kl}\rightarrow
\beta ^p_{kl}+\nabla_k u_l^p- \epsilon _{kl} \omega ^p_3$,\,
$
\phi^p_{l}\rightarrow
\phi^p_{l}+\partial _l \omega ^p_3$.
The general defect has
a displacement field
\vspace{-1.8em}~\\
\begin{eqnarray}
u_l=- \delta (V_2)[b_l- \Omega  \epsilon _{3lr}(x_r-\bar x_r)].
\label{@}\end{eqnarray}
\vspace{-1.8em}~\\
The
first
term is a dislocation, the second term  a disclination.
According to Fig.~\ref{stack}, the latter can be read as a superposition of dislocations with
the same
Burgers vector $\tilde b_l=-\int_{ \bar x}^x dx'_r\Omega  \epsilon _{3lr}$.
The former may be viewed
 as a dipole of disclinations:
$
-\bar\nabla_l[
-\sfrac{1}{2} b_m\epsilon _{3km}]
\epsilon _{3kr}(x_r-\bar x_r)$.

{\bf 6.}
Let us now derive the emerging
 gravitational forces
in the world crystal.
Consider the partition function, at unit temperature,
of the world crystal 
which we take to be three-dimensional, for simplicity:
\begin{eqnarray}
&&Z=\sum_{n_{ij}({\bf x})}\prod_{{\sbf x},i}\left[\frac{du_i({\sbf
        x})}{a}
\right]
e^{-H}.
\label{@}\end{eqnarray}
\vspace{-1.2em}~\\
In linear elasticity,
the energy depends quadratically only on the
difference between the elastic and the plastic strain tensors
$u_{ij}=\sfrac{1}2(\nabla_i u_j+\nabla_ju_i)$
and $u_{ij}^p$, and reads on the lattice
\begin{eqnarray}
H=\frac
{\mu}4\sum_{{\sbf x}}\sum_{i<j}[
\nabla_iu_j({\bf x})
\!+\!\nabla_ju_i({\bf x})
\!-\!n_{ij}({\bf x})]^2
.\label{@}\end{eqnarray}
\vspace{-1.3em}~\\
Here $\mu$ is the elastic constant \cite{RE1}, and
the integer numbers $n_{ij}$
of are the lattice
versions of $2u_{ij}^p$
 in Eq.~(\ref{@def}).
This partition function explains
for low temperature
the correct
classical specific heat. If the temperature is increased, it reaches
a point where
the configuration entropy of the defects wins over the
Boltzmann factors of their
energy,
and the world crystal melts.

We have shown in \cite{MVF}
that in order to arrive at the proper Newton forces
at long distances
we have to insert one more derivative
in the lattice action
and start out with what is
 called  the {\em floppy world crystal\/}
where
\begin{eqnarray}
H\!= \!
{\mu}\sum_{{\sbf x}}\!\sum_{i<j}\!\left\{\nabla_k[
\nabla_iu_j({\bf x})
\!+\!\nabla_ju_i({\bf x})
\!-\!n_{ij}({\bf x})]\right\}^2 .
\label{@}\end{eqnarray}
\vspace{-2.5em}~\\

The partition function
depends on the defect configuration
only via the defect tensor
formed from $n_{ij}$, i.e.,
on 
 $ \eta
_{ij}$.
It is a functional of this tensor
which can be expanded into powers
of
$ \eta _{i}{}^i,\,
\eta _{ij} \eta ^{ij},\,
\eta _{ij}
\eta ^{jk}
\eta _{k}{}^i,\,
\eta _{ij}
\eta ^{ij}
\eta _{k}{}^k,
\, \dots~$.
The expansion coefficient are proportional
to powers of the Planck length, for each $\eta_{ij}$ two powers.
Since $ \eta _{ij}$
is the
defect represention of the Einstein tensor $G_{ij}$,
the partition function defines
a gravitational action
which is a power series of
the Einstein tensor.
To leading (second) order in the Planck length
is is proportional 
to the scalar $G=G_i{}^i=-R$.

Note that the gravitational action
arises in this model
from the {\em entropy\/}
of the fluctuations \cite{ENTR}
in the same way as rubber elasticity
in polymer physics \cite{PIP}.

The
defect tensor
$ \eta _{ij}\hat{=} G_{ij}$
can be decomposed into
an Cartan
part and a torsion part
as in Eq.~(\ref{@dc}).
From the equivalence
of defects illustrated in Fig.~\ref{stack}
it is now obvious that we
can re-express the
action,
which contains only to defect tensor $ \eta _{ij}\hat{=} G_{ij}$,
completely in terms of the dislocation density $ \alpha _{ij}$, i.e.,
in terms of the torsion tensor $S_{lkj}$ via Eq.~(\ref{@TOR}).
The Cartan curvature tensor is
then identically zero, 
showing that
Einstein's teleparallel formulation of gravity
is completely equivalent
to the original Einstein theory.
Alternatively, we may make the torsion vanish identically,
and recover
the original
Einstein theory.

In addition, there exists
an infinite number of intermediate formulations
of the theory with both Riemann-Cartan curvature and
torsion in some well-defined mixture.

{\bf 7.}
Generalizing
the defect relations
(\ref{@dc}) and
(\ref{@2def})
to $D\geq 4$ spacetime dimensions
and allowing for
large deviations
from Euclidean space, we find \cite{KONDOm}
\vspace{-1.6em}~\\
\begin{eqnarray}
    G_{\mu \nu }= G^{\rm C}_{\mu \nu } - \sfrac{1}{2}
	D^*{}^\lambda  \left( S_{\mu \nu ,\lambda }- S_{\nu \lambda ,\mu }
	+ S_{\lambda \mu ,\nu } \right)
\label{3.51}\end{eqnarray}
\vspace{-1.6em}~\\
where $G_{\mu \nu }$ is the Einstein tensor
and $G^{\rm C}_{\mu \nu }$ its Cartan
version, while
$S_{\mu \kappa }{}^{,\tau} $
is the
Palatini  tensor
related to the torsion field
$S_{\mu \kappa }{}^{\tau} $
by
 \begin{eqnarray}
   \sfrac{1}{2} S_{\mu \kappa } {}^{,\tau} \equiv S_{\mu \kappa }{}^\tau
	  + \delta _\mu  {}^\tau S_\kappa{}_ \lambda {}^ \lambda  -
\delta _\kappa {}^\tau
	     S_\mu{}_ \lambda {}^ \lambda.
\label{3.40}
\label{palat@}
\end{eqnarray}
\vspace{-1.4em}~\\
The symbol $D_\mu$
denotes the covariant
derivative defined by
$D_\mu v_ \nu \equiv
\partial _\mu v_ \nu - \Gamma _{\mu \nu }{}^ \lambda v_ \lambda
 $,~
$D_\mu v^ \lambda \equiv
\partial _\mu v^  \lambda  + \Gamma _{\mu \nu }{}^  \lambda   v^ \nu
 $, and $D^*_\mu\equiv
D_\mu+2 S_{\mu \kappa }{}^ \kappa$.
The defect conservation laws
(\ref{@CL}) read
\vspace{-1.7em}~\\
\begin{eqnarray}&&
D_\mu ^*G^{\rm C}_{\, \,\lambda }{}^{\mu  }+
2 S^{ \nu  \lambda}{}^{ \kappa} G^{\rm C}_{\,\kappa \nu} -\sfrac{1}{2}S^
{\nu \kappa,\mu}
R^{{\rm C}}_{\, \lambda\mu \nu \kappa}=0\,,
\label{@DERG}\\
&&~~~~~~~~~D^*{}^\mu S_ {\lambda  \kappa ,\mu}=
G^{\rm C}_{\, \lambda    \kappa }-
G^{\rm C}_{\, \kappa  \lambda   }.
\label{@DERG2}\end{eqnarray}
\vspace{-2.em}~\\
They are Bianchi identities
ensuring the single-valuedness of
observables, connection
$
 \Gamma _{\mu \nu }{}^ \lambda
 $  and metric $g_{\mu \nu }$,
via the integrability conditions
$[\partial _ \sigma ,\partial _\tau ]
 \Gamma _{\mu \nu }{}^ \lambda=0$ and
$[\partial _ \sigma ,\partial _\tau ]
 g _{\mu \nu }=0$.

In a four-dimensional
Riemann-Cartan spacetime,
the geometry is
described by
the
direct
 generalizations of
{\em translational\/} and {\em rotational\/} defect  gauge fields
$h_{ij}$ and
$
A_{ijk}$,
which are here
the vierbein
 field $h^\alpha{}_\mu$,
and the spin connection
$
A_{\mu \alpha }{}^ \beta $.
The square of the former is
the  metric $g_{\mu \nu }=h^ \alpha {}_\mu h_ \alpha {}_ \nu $.
The latter is defined by
the covariant derivative
$
D_ \lambda  h_\beta {}^\mu =\partial _ \lambda h_\beta {}^\mu
-A_{ \lambda  \beta }{}^  \gamma  h_ \gamma{}^\mu
+ \Gamma _{ \lambda \nu}{}^ \mu h_ \beta {}^ \nu\equiv
D^L_ \lambda  h_\beta {}^\mu
+ \Gamma _{ \lambda \nu}{}^ \mu h_ \beta {}^ \nu.$
The field strength
of $A_{\mu \alpha }{}^ \beta
\equiv (A_\mu)_ \alpha {}^ \beta $
\vspace{-1.9em}~\\
\begin{eqnarray}
%$
F_{\mu \nu  \beta }{}^  \gamma \equiv
\{\partial _\mu A_ \nu
-\partial _\nu A_ \nu -[A_\mu,A_ \nu ]\}_ \lambda{} ^ \kappa,
%$
\label{@FS}\end{eqnarray}
\vspace{-1.9em}~\\
determines the Cartan curvature
$R^{{\rm C}}_{\mu \nu  \lambda }{}^ \kappa
 \equiv h^ \beta {}_ \lambda  h_ \gamma {}^ \kappa
F_{\mu \nu   \beta }{}^ \gamma  $.
The
field strength of $h^ \gamma  {}_\nu$
is the
torsion:
\vspace{-1.9em}~\\
\begin{eqnarray}
%$
S_{ \alpha  \beta }{}^ \gamma \equiv \sfrac{1}{2}h_ \alpha {}^\mu h_ \beta {}^ \nu
[D^L_\mu h^ \gamma {}_ \nu -(\mu\leftrightarrow  \nu )].
%$
\label{@FS}\end{eqnarray}
\vspace{-1.9em}~\\
The relations (\ref{3.51}),
(\ref{@DERG}), and
(\ref{@DERG2})
follow from this.

{\bf 8.}
The
theory is gauge invariant
under local Lorentz transformations
as a direct consequence
of the fact that the metric can alternatively be written as
\vspace{-1.9em}~\\
\begin{eqnarray}
g_{\mu \nu }=    h^   \gamma   {} _\mu
 \Lambda ^a{}_   \gamma
 \Lambda _a{}^  \beta
 h_ \beta  {}_ \nu,
\label{@GDEC}\end{eqnarray}
\vspace{-1.9em}~\\
where
$
 \Lambda _a{}^  \beta$ is an {\em arbitrary  local\/} Lorentz transformation,
and that the Einstein-Hilbert Lagrangian
${\cal L}_{\rm EH}=-(1/2 \kappa ) R$
is
{\em independent\/}
of $ \Lambda ^a{}_ \alpha $.
The extra
$ \Lambda _a {}^  \beta $
transforms the gauge field
$
A_{\mu \alpha }{}^ \beta$ as
\vspace{-1.9em}~\\
\begin{eqnarray}
A_{\mu \alpha }{}^ \beta
\rightarrow
A_{\mu \alpha }{}^ \beta+
 \Delta A_{\mu \alpha }{}^ \beta,~~~
 \Delta A_{\mu \alpha }{}^ \beta\equiv  \Lambda _a{}^ \beta \partial _\mu
 \Lambda ^a{}_ \alpha .
\label{@GDE}\end{eqnarray}
\vspace{-1.9em}~\\[-1.3em]

At this point we are ready to introduce the  {\em new\/} gauge invariance
announced in the title:
we allow
$\Lambda _a{}^  \beta$
in Eq.~(\ref{@GDEC})
to be
a {\em multivalued Lorentz transformation\/}.
This is {\em not\/} integrable, so that
$ \Delta A_{\mu \alpha }{}^ \beta%\equiv  \Lambda _a{}^ \beta \partial _\mu \Lambda^a{}_ \alpha
$ is a {\em nontrivial\/} gauge field.
Indeed, the rotational field strength
$
F_{\mu \nu \alpha }{}^ \gamma $
can be expressed as
 $
F_{\mu \nu \alpha }{}^ \gamma \equiv
 \Lambda _ a {}^ \gamma [\partial _\mu,\partial _ \nu ]
 \Lambda ^a{}_ \alpha \neq0$
and yields a nonzero Cartan curvature
$
R^{{\rm C}}_{\mu \nu  \lambda }{}^ \kappa
  \neq0$.
The important observation
is that a {\em multivalued\/} $ \Lambda^a{}_ \alpha $ is
able to {\em change the geometry\/}
 \cite{MVFr}.
The right-hand side of
(\ref{3.51}) is {\em independent\/} of the vector field $
A_{\mu \alpha }{}^ \beta$.
This allows us
to shuffle  torsion into Cartan curvature and back,
fully or partially, by complete
analogy with the defect
transformations in
two-dimensional crystals
 in Fig.~\ref{stack}.
We may  choose
for $ A_{\mu \alpha }{}^ \beta$
any function we like.
For example we may choose 
it to
make the torsion
vanish,
and
$  A_{\mu \alpha }{}^ \beta $
reduces to the
usual
spin connection of Einstein's gravity, the
well-known combination of the objects of 
anholonomity
~\\[-1.7em]
\begin{equation}
 \Omega _{\mu  \nu }{}^ \lambda
 =\sfrac{1}{2}[
h_ \alpha {}^ \lambda \partial _\mu h^ \alpha {}_ \nu
-(\mu\leftrightarrow \nu )].
\end{equation}
~\\[-1.7em]
In the opposite extreme
 $ A_{\mu \alpha }{}^ \beta=0$,
the Cartan curvature is zero, spacetime is teleparallel,
and
 the Lagrangian
is equal to the combination of torsion tensors:
~\\[-1.7em]
\begin{equation}
{\cal L}_S\!=\!-\frac1{2 \kappa }(
-4D_\mu S^\mu+S_{\mu \nu  \lambda }S^{\mu \nu  \lambda }+2
S_{\mu \nu  \lambda }S^{\mu \lambda  \nu }-4 S^\mu{}S_\mu ),
\end{equation}
~\\[-1.7em]
where $S_\mu\equiv S_{\mu \nu }{}^ \nu $.

In any of the new gauges,
the correct gravitational field equations are derived 
by extremizing
the Einstein-Hilbert action
% in terms of $h_ \alpha {}^\mu$ and $A_{\mu \alpha }{}^ \beta $ as:
%
\vspace{-1.9em}~\\
\begin{equation}
{\cal A}_{\rm EH}=
-\frac1{2\kappa }
\int d⁴x \sqrt{g}\,R^{{\rm C}}
+\int d⁴x \sqrt{g}\, {\cal L}_S
+{\cal A}_{\rm GF}
, 
\label{@}\end{equation}
\vspace{-1.5em}~\\
where
${\cal A}_{\rm GF}$ is a
functional 
of $h^\alpha{}_\mu$ and $ A_{\mu \alpha }{}^ \beta$
fixing some 
convenient gauge.
For ${\cal A}_{\rm GF}=\delta[ A_{\mu \alpha }{}^ \beta]$
this leads to the teleparallel theory,
and for  ${\cal A}_{\rm GF}=\delta[ S_{\alpha,\beta, \gamma }[
h^\alpha{}_\mu, A_{\mu \alpha }{}^ \beta]]$
we re-obtain Einstein's original theory.

{\bf 9.}
Adding
 matter fields of masses $m$
to the Einstein Lagrangian,
and varying with respect to
$h^ \alpha {}_\mu$,
%and $A_{\mu \alpha} {}^ \beta $,
we find in the zero-torsion gauge
the Einstein equation
\vspace{-1.7em}~\\
\begin{equation}
G_{\mu \nu }= \kappa T_{\mu \nu },
\label{@}\end{equation}
\vspace{-1.7em}~\\
where
$T_{\mu \nu }$ is the sum over the
symmetric energy-momentum tensors of all matter fields.
%${\!\!\phantom{I}}^{\rm m}T_{\mu \nu }$.
Each contains the canonical energy-momentum tensor
$\!\stackrel{\rm m}{\phantom{r}}\hspace{-6pt}\Theta_{\mu \nu }$
and the spin current densities $\!\stackrel{\rm m}{\phantom{r}}\hspace{-3pt}\Sigma_{\mu \nu}{}^{\!,\lambda}$
in the
combination due to Belinfante \cite{BELIN},
\vspace{-1.2em}~\\\!\!\!
       \begin{equation} 
   \stackrel{\rm m}{\phantom{r}}\hspace{-3pt}{T}_{ \kappa  \nu}  \,
 \!\!=\,\stackrel{\rm m}{\phantom{r}}\hspace{-3pt}{\Theta}_{ \kappa  \nu }
       -\!\sfrac{1}{2} D^*{}^\mu  \left(
    \stackrel{\rm m}{\phantom{r}}\hspace{-3pt}{\Sigma}_{ \kappa  \nu }
      {}_{,\mu }-\!
    \stackrel{\rm m}{\phantom{r}}\hspace{-3pt}{\Sigma}_{  \nu \mu,\kappa  } +
\!    \stackrel{\rm m}{\phantom{r}}\hspace{-3pt}{\Sigma}
      _{\mu \kappa ,~ \nu }\right)\!,  \!\!
\label{4.83}\end{equation}
\vspace{-1.2em}~\\
which is the matter analog
of
the defect relation
(\ref{3.51}).

The new gauge invariance of (\ref{3.51})
has the
physical consequence that
the external gravitational field
in the far-zone of a celestial body
does not care whether
angular momentum comes from rotation of matter or from internal spins.
The off-diagonal elements
of the metric in the far-zone, and thus the Lense-Thirring effect
measured in \cite{STAN},
depend only on the total
angular momentum
$J^{ \lambda \mu } =
\int d^3x(
x^ \lambda  T^\mu{}^ 0
-x^ \mu  T^ \lambda {}^ 0)$, which
by the Belinfante relation (\ref{4.83})
is the
 sum
of orbital
 angular momentum
$L^{ \lambda \mu } =
\int d^3x(
x^ \lambda  \Theta^\mu{}^ 0
-x^ \mu  \Theta^ \lambda {}^ 0)$
and spin
$S^{ \lambda \mu}=\int d^3x  \Sigma ^{ \lambda \mu,0}$.
A
star consisting of polarized
matter
has the same
external gravitational field in the far-zone
as a star rotating with the corresponding
orbital angular momentum.        This
is
the {\em universality of
orbital momentum and intrinsic
angular momentum\/}
in gravitational
physics observed in Ref.~\cite{OA}.

Since torsion is merely a
{\em new-gauge degree\/} of freedom
in describing a gravitational field, it cannot be detected experimentally, not even
by spinning particles.
A
field with arbitrary spin may
be 
 coupled to gravity via the covariant derivative
$D_\mu\equiv  \partial _\mu {\bf 1}+
\sfrac{i}{2}
A_{\mu  \alpha }{}^ \beta
\Sigma^\alpha{}_\beta
 $, 
where
$\Sigma^\alpha{}_\beta
$ are the generators 
of the Lorentz 
group, in the Dirac case
$\Sigma_\alpha{}_\beta
=
\frac{i}{4}[ \gamma ^ \alpha , \gamma _ \beta ]
$. But since 
the torsion is a tensor,
we may equally well use an infinity
of alternative
covariant derivatives
$ D^q_\mu\equiv  \partial _\mu {\bf 1}+
\sfrac{i}{2}
 A^q_{\mu  \alpha }{}^ \beta
\Sigma^\alpha{}_\beta
$, 
where 
$A^q_{\mu  \alpha }{}^ \beta
\equiv
 A_{\mu  \alpha }{}^ \beta-gK_{\mu  \alpha }{}^ \beta
$, and $K_{\mu\alpha\beta}=
h_\alpha{}^\nu
h_\beta{}^\lambda
K_{\mu \nu\lambda}\equiv
h_\alpha{}^\nu
h_\beta{}^\lambda(
S_{\mu\nu\lambda}
-S_{\nu\lambda\mu}
+S_{\lambda\mu\nu}
)$.
Any 
coupling constant $q$ 
is permitted by covariance.
In order to see which $q$ is physically correct
we come back to the above-discussed 
photon mass problem, and 
consider the 
covariant 
electromagnetic field tensor
$F_{\mu\nu}^q
\equiv
 D_\mu^q A_\nu- D_\nu^q A_\mu$.
Working out the covariant derivative
we find
$ \partial_\mu A_\nu-\partial_\nu A_\mu-2(1-q)S_{\mu\nu}{}^\lambda
A_\lambda$, which shows
that 
Maxwell
Lagrangian 
$
-\sfrac{1}{4}F^q_{\mu\nu}F^q{\,} ^ {\mu\nu}$
acquires a mass term, 
unless we fix the coupling 
strength to the value $q=1$.

For this value of $q$,
a little algebra 
\cite{GFCM,MVF}
shows that
the torsion drops out from 
the gauge field
$ A^q_{\mu  \alpha }{}^ \beta $.
This
reduces to the 
good-old Fock-Ivanenko
spin connection that has been
used in Einstein gravity
without torsion:
\begin{eqnarray}
 A^1_{\mu  \alpha }{}^ \beta = \bar
A_{\mu  \alpha }{}^ \beta=h_\alpha{}^\nu h^\beta{}^\lambda(
\Omega_{\mu\nu\lambda}
-
\Omega_{\nu\lambda\mu}
+\Omega_{\lambda\mu\nu}
).
  \label{@AHO}\end{eqnarray}

Having ensured  
that the photon does
not couple to torsion,
we must also prevent all 
all other spinning baryonic matter
to do so, 
to avois 
giving a mass
to the photons
via virtual processes
of the type discussed above an illustrated in Fig.~\ref{@baryon}.

{\bf 10.}
How about the motion of a 
spinless point particle in the infinitely
many different
descriptions of the same theory?
Since the metric
$
g_{\mu \nu }=    h^   \gamma   {} _\mu
 \Lambda ^a{}_   \gamma
 \Lambda _a{}^  \beta
 h_ \beta  {}_ \nu
 $
is independent of the local Lorentz transformations
$ \Lambda ^a{} _\alpha $, and the action ${\cal A}=-mc\int
ds=-mc\int({g_{\mu \nu }dx^\mu dx^ \nu })^{1/2} $
depends only on $g_{\mu \nu }$,
the
trajectories are
geodesics for all $ \Lambda _a{}^  \beta$.
The same result can of course be obtained 
my integrating the
local conservation law 
of the total energy-momentum tensor
$
T_{\mu\nu}$
along a thin world-tube.

A spinning particle
``sees" 
the gauge field of Lorentz transformations
$A^q_{\mu\alpha}{}^\beta$, but it
 does so only via the $q=1$-version
(\ref{@AHO}). This contains 
only 
the vierbein fields, not the torsion, and 
is invariant
under the multivalued 
version of the
gauge transformation
(\ref{@GDE}).
Hence
the motion  
of a spinning particle 
is blind  the torsion 
field, which can
therefore not be detected by any experiment.

{\bf 11.}
What we have done can be understood 
better by a simple analogy.
Insted of Einstein's theory, we
consider a 
model of a real field 
$\rho$ with an Euclidean Lagrangian
${\cal L}=(\partial_\mu \rho)^2-\rho^2+\rho^4$ and a 
partition function
$ Z=\int {\cal D}\psi {\cal D}\psi^* \,e^{-\int dx {\cal L}}
$. The field $\rho$ is the analog of the metric $g_{ \mu\nu}$.
We may now trivially 
introduce an extra gauge structure
by re-expressing
the Lagrangian in terms of 
a complex field $\psi= e^{i \theta}\rho$
and a gauge field $A_\mu$ as
${\bar{\cal L}}=|(\partial_\mu -iA_\mu)\psi|^2-|\psi|^2+|\psi|^4$.
Now we form the partition function
$\bar Z=\int {\cal D}\psi {\cal D}\psi^*{\cal D}A_\mu \Phi\,e^{-\int dx \bar{\cal L}}
$,
where $\Phi$ is an arbitrary 
gauge-fixing functional
multiplied by the associated Faddev-Popov determinant.
The new $\bar Z$
is completely 
equivalent to the original 
$ Z$.
Obviously there is no way of observing $A_\mu$.
The partition function $Z$ plays the role of Einstein's theory,
whereas $\bar Z$ gives
its reformulation in 
terms of a gauge field, which does not change the physical content of
the theory.
The decomposition $\rho=\psi^*\psi
=(\rho e^{-i\theta})
(e^{i\theta}\rho)$ is the analog of 
the decomposition ~(\ref{@GDEC}).

{\bf 12.}
Higher gradient terms in elastic energy
of the world crystal will generate
an extra action 
${\cal A}_{A_\mu}$ of the
gauge field $A_\mu{}_ \alpha {}^ \beta$ \cite{HO}.
 This would, in general,  violate the 
new symmetry discussed above
and give torsion a life of its own.
However, 
as long as the gravitational effects 
of spinning constituents 
in celestial bodies are
suppressed with respect to that
of
the orbital angular momenta
by many orders of magnitude,
there is not much sense in conjecturing
explicit forms of 
${\cal A}_{A_\mu}$, unless we want to compete with
string theory in setting up an ultimate {\em theory of everything\/}
as a substitute of religion.

{\bf 13.}
In summary, we have shown
that
if the Einstein-Hilbert Lagrangian
is expressed in terms of the
translational and rotational gauge fields
$h^ \alpha {}_\mu$ and  $A_\mu{}_ \alpha {}^ \beta$,
the  Cartan curvature can be converted
to torsion
and back, totally or partially,  by a
{\em new type of multivalued
gauge transformation\/}
in Riemann-Cartan spacetime, a {\em hypergauge transformation}.
In this
general formulation,
Einstein's original theory is obtained by
going to the
zero-torsion hypergauge, while
his
teleparallel theory is
in the hypergauge in which the Cartan
curvature tensor vanishes.
But any intermediate choice of the field
 $A_\mu{}_ \alpha {}^ \beta$
is also allowed.
~\\~\\[-.2em]
Acknowledgment:
I thank F.W. Hehl, J.G. Pereira, and especially
Jan Zaanen
for a critical reading of the manuscript.

%\vspace{-2em}


\begin{thebibliography}{99}
\bibitem{RE1}
We ignore here the second
elastic constant since it is irrelevant to the argument.
\end{thebibliography}
\end{document}